\documentclass[a4paper,fleqn,usenatbib]{mnras}
\usepackage{graphics}
\usepackage{graphicx}
\usepackage{psfig}
\input{epsf}
\usepackage{bm}
\usepackage{amsmath}
\usepackage{amssymb}
\usepackage{url}
\usepackage[T1]{fontenc}
\usepackage{ae,aecompl}
\usepackage{widetext}
\usepackage{multirow}

\title[Intrinsic alignments super-sample covariance]{Cosmological consequences of intrinsic alignments super-sample covariance}
\author[Saeed Ansarifard \& S. M. S. Movahed]{
Saeed Ansarifard$^{1}$\thanks{E-mail: s\_ansarifard@sbu.ac.ir} \& S. M. S. Movahed$^{1}$\thanks{E-mail: m.s.movahed@ipm.ir}
\\
$^{1}$Department of Physics, Shahid Beheshti University,  1983969411, Tehran, Iran}

\begin{document}
\maketitle

\begin{abstract}
We examine cosmological constraints from high precision weak lensing surveys including super-sample covariance (SSC) due to the finite survey volume. Specifically, we focus on the contribution of {\it beat coupling} in the intrinsic alignments as a part of full cosmic shear signal under flat sky approximation. The SSC-effect grows by going to lower redshift bin and indicates considerable footprint on the  intermediate and high multipoles for cumulative signal to noise ratio (SNR).  The SNR is reduced by $\approx 10\%$ as a consequence of including the intrinsic alignments super-sample covariance, for the full cosmic shear signal, depending on the amplitude of intrinsic alignments, the ellipticity dispersion and the survey redshift ranges, while the contribution of photometric redshift error can be ignored in the cumulative signal to noise ratio. Using the Fisher-matrix formalism, we find that the impact of large modes beyond the volume of the surveys on the small modes alters the intrinsic alignments. However, corresponding impact on the cosmological parameters estimation is marginal compared to that of for gravitational weak lensing, particularly, when all available redshift bins are considered.  Our results also demonstrate that, including SSC-effect on the intrinsic alignments in the analytical covariance matrix of full  cosmic shear leads to increase marginally the confidence interval for $\sigma_8$ by $\approx 10\%$ for a sample with almost high intrinsic alignments amplitude.

\end{abstract}

\begin{keywords}

cosmological parameters - methods: numerical - gravitational lensing: weak - large-scale structure of Universe

\end{keywords}

\section{Introduction}
Light deflection on gravitational potentials sourced by the cosmic matter distribution is referred to weak gravitational lensing (WL) \citep{bartelmann2001weak}. This phenomenon has received considerable attention \citep{schneider1996detection,rhodes2001detection,hoekstra2002measurement,kaiser2000large,van2000detection,bacon2000detection,bacon2003joint,brown2003shear}.
To put observational constraints on the growth of cosmic structure growth and to asses the properties of dark energy in more details, many projects have been proposed focusing on the cosmic shear measurements such as the Dark Energy Survey \footnote{\texttt{http://www.darkenergysurvey.org}} (DES: \cite{abbott2018dark,troxel2018dark}), the Kilo-Degree Survey \footnote{\texttt{http://kids.strw.leidenuniv.nl}} (KiDS: \cite{hildebrandt2017kids,kohlinger2017kids}). A particular interest on the weak lensing cosmology has been devoted to the Subaru Hyper Suprime-Cam survey\footnote{\texttt{http://hsc.mtk.nao.ac.jp/ssp/}} \citep{aihara2018hyper,hikage2019cosmology}, the Rubin Observatory Legacy Survey of Space and Time (LSST) \footnote{\texttt{http://www.lsst.org}} \citep{2019ApJ...873..111I}, the ESA Euclid mission\footnote{\texttt{http://sci.esa.int/euclid/}} \citep{2011arXiv1110.3193L,amendola2018cosmology} and the NASA Nancy Green Roman Space Telescope mission \footnote{\texttt{https://roman.gsfc.nasa.gov/}},  (see also \citep{huff2014seeing,jee2016cosmic,dore2018wfirst}). In the limit of small distortions of the galaxy shapes, the cosmic shear ($\gamma^{\rm G}$)  is able to map out the matter distribution in the Universe without any assumptions on its dynamical state \citep{blandford1991distortion,miralda1991correlation,kaiser1992weak}.

Combining the results of cosmic shear and galaxy clustering improves the constraints on the value and evolution of dark matter density \citep{joudaki2017kids,van2018kids+,yoon2019constraints,abbott2018dark}. Interestingly, it is possible to confine the value of dark energy density by the cosmic shear accomplished by  baryonic acoustic oscillation (BAO) and supernova probes derived from photometry surveys \citep{hoekstra2008weak,merkel2017parameter,Abbott2019cosmological,omori2019dark}. More recently, combinations of redshift-space distortions, cosmic shear and CMB observations have been used to examine the general relativity \citep{amon2018kids+,Ferte2019testing}.

Utilizing various surveys with different geometries and pipelines
leads to different results in the high precision regime not only for
the best fit values of cosmological parameters, but also for the
associated confidence intervals \citep{chang2018unified}. Therefore,
it is crucial to control various discrepancies ranging from the
source of errors, contaminations such as catalog shape, point spread
function models, photometric redshift estimation to intrinsic
alignments (IA) in the designing future high accuracy experiments
\citep{mandelbaum2018weak}. 

The IA of galaxies especially constitutes an inevitable contribution on the measured cosmic shear signal \citep{croft2000weak,heavens2000intrinsic}. Therefore, what we measure in the observations is a superposition of intrinsic correlations of neighboring galaxies and those correlations induced by cosmic lenses which is called full cosmic shear signal, $\gamma^{\rm I+G}=\gamma^{\rm G}+\gamma^{\rm I}$ \citep{hirata2004intrinsic}. The cosmological tidal field can explain the existence of IA shear ($\gamma^{\rm I}$) in the cosmic shear measurements. This excess value as a systematic error is a direct consequence of tidal torques and forces affecting on the spin of blue spiral galaxies and stretch the red elliptical galaxies, respectively \citep{catelan2001intrinsic,crittenden2001spin,schaefer2009galactic}.  Subsequently, in the presence of intrinsic ellipticity, the two-point correlation function of full cosmic shear signal, is essentially affected by additional terms which has to be taken into account for more robust and high precision observational constraints \citep{hirata2004intrinsic,krause2015impact}.

Many efforts have been made to control or mitigate the $\gamma^{\rm I}$ in the cosmic shear probes \citep{joachimi2015galaxy,troxel2015intrinsic}. Meanwhile, some researches have been concentrated on the separation of IA contribution from WL in the ellipticity signal of weak lensing surveys \citep{blazek2012separating,schafer2017angular,merkel2017imitating}.

It is worth noting that, the IA itself has considerable impact on the better understanding of distribution and evolution of matter in the universe  \citep{chisari2013cosmological}. The IA depends on the morphology of the galaxies as well as on their location inside the larger structures representing the complexity of IA effect in the statistical intrinsic ellipticity probes \citep{georgiou2019dependence}. According to the most recent measurements, the $\gamma^{\rm I}$-signal of spiral galaxies, thought to be related to the angular momentum direction, can be ignored almost entirely, while intrinsic shape correlations of elliptical galaxies, which traces back to shearing by tidal forces, are experimentally confirmed. They typically constitute the amount to 10\% of the lensing signal over a wide multipole range \citep{samuroff2019dark,johnston2019kids+}. In general, this intrinsic ellipticity correlation $\langle \gamma^{\rm I} \gamma^{\rm I} \rangle$, if modeled through tidal interaction, depends on the matter spectrum, either in the linear or nonlinear approximation \citep{blazek2019beyond}. There are uncertainties in the large-scale structure observations due to the finite volume of cosmological surveys, and this is in particular true for weak lensing. Typically, we encounter with a systematic error in estimating spectra called super sample-covariance (SSC) \citep{hamilton2006measuring}. Such effect is usually known as the {\it beat coupling} due to the impact of large modes beyond the volume of surveys on the small modes. This phenomenon produces a systematic error on the matter power estimator \citep{rimes2006information,takada2013power}.

It has been confirmed that, the SSC decreases the constraining capability of many different observations such as BAO, Alcock-Paczynski (AP) and redshift-space distortion (RSD) \citep{li2018galaxy}.  Combining the cluster number counts measurement with WL almost diminishes the footprint of SSC in cosmological parameter constraints \citep{takada2014joint}. Some numerical algorithms have been proposed to improve the SSC estimation on the power spectrum \citep{lacasa2018super}. It has been also suggested to consider the zero mode fluctuation as a new free parameter and to carry out observational constraints on the corresponding parameters \citep{li2014super}.  More recently, a mitigation strategy for the reconstruction of SSC has been proposed resulting in alleviating error volumes in the joint analysis \citep{digman2019forecasting}.

Many surveys construct the covariance matrix from simulations \citep{kohlinger2015direct,kohlinger2017kids}, but in our investigation, we work with an analytical approximation \citep{eifler2009dependence} which allows us to incorporate the effects of super sample-covariance clearly. Specifically, there are known terms corresponding to the SSC-effect on the cosmic shear, but the analogous effect on intrinsic alignments is appreciable in particular when one aims to reach percent-level precision.

In this paper, we rely on the $\gamma^{\rm I}$ as a part of full cosmic shear signal in the tomographic surveys and include the SSC term  in the full analytical covariance matrix used for Likelihood analysis. The SSC term in this work is computed by the {\it beat coupling} in the weakly non-linear regime.  We examine the effect of SSC term in the covariance matrix in order to find confidence intervals of  cosmological parameters including dark energy equation of state ($w_0$), density of dark matter ($\Omega_m$), the variance of the linear density contrast on the scale $R_8=8h^{-1}$ Mpc ($\sigma_8$), scalar spectral index ($n_s$) and the present value of Hubble parameter ($H_0$). To realize the SSC-effect in detail, we also consider the variation of redshift bin size evaluated by  photometric redshift error ($\sigma_z$) and associated redshift values. The impact of shot noise will also be considered. In addition, the competition between the contribution of IA and WL in the SSC will be studied. To this purpose, we utilize Fisher Matrix Formalism \citep{1997ApJ...480...22T,bassett2011fisher}.

The outline of this paper is as follows:  In section \ref{sec:method}, we present theoretical requirements for computing the IA and full cosmic shear signal with and without considering SSC effect. The results for Fisher forecast are described in section \ref{sec:results}. Section \ref{sec:summary} will be devoted to summary and conclusions.
Throughout this paper, we assume a spatially flat $\Lambda \rm CDM$ model and the value of corresponding parameters are $\{\Theta\}:\{H_0 = 72.0, \Omega_m = 0.300, \Omega_b = 0.0424, \sigma_8 = 0.800, n_s = 0.940, w_0 = -1\}$.

\section{Method}\label{sec:method}
In this section, we introduce our mathematical notion for examining the SSC-effect on the IA and more generally full cosmic shear signals. The WL and IA shears are usually written in the real space in the form of line of sight integral:
\begin{equation}\label{eq:1}
\gamma^{\rm A}_{\theta}(i) = \int \mathrm{d}\chi W_i^{\rm A} \gamma^{\rm A}\left(\bold{x},z\right)
\end{equation}
the superscript "${\rm A}$" is replaced by either ${\rm G}$ or ${\rm I}$  and  the subscript "$i$" shows the $i$-{th} redshift bin, $\chi$ is comoving distance, $\bold{x} = \left(\chi\theta,\chi \right)$ is 3-dimension (3D) positional vector and $\gamma^{\rm A}\left(\bold{x},z\right)$ is the 3D shear field expressed by matter density contrast.
The kernel functions in Eq. (\ref{eq:1}) read as \citep{johnston2019kids+,merkel2017parameter}:
\begin{eqnarray}\label{eq:kernel}
    W_i^{\rm G} \left( \chi \right) &\equiv& \chi (1+z\left(\chi\right))\int_{\chi}^{\chi_{\mathrm{max}}} \mathrm{d}\chi^\prime  \frac{\chi^\prime - \chi}{\chi^\prime} n_i(\chi^\prime) \nonumber\\
    W_i^{\rm I} \left( \chi \right)  &\equiv& - \frac{A_{\rm I} n_i(\chi)}{100{H_0}^{2} }
\end{eqnarray}
here $A_{\rm I} $ is simply a dimensionless free parameter which is order of unity. We choose $A_{\rm I} = 1$ and $A_{\rm I} = 3$ for samples with normal and excess intrinsic alignment. The factor 100 is chosen in such a way to match with a value obtained from the SuperCOSMOS Sky Survey \citep{johnston2019kids+,samuroff2019dark,merkel2017parameter,brown2002measurement}. The normalized galaxy distribution, $n_i(\chi)$, depends on survey characteristics \citep{bartelmann2001weak}:
\begin{equation}\label{eq:nz}
    n(z) = \frac{q_0z^2}{z_0^2} e^{ -\left(\frac{z}{z_0}\right)^\beta}
\end{equation}
where $z_0=0.64$, $\beta = 1.5$ and $q_0\equiv\beta/z_0$ and here we consider 10 redshift bins. The bins' boundaries are chosen in such a way that the number of galaxies in each bin to be identical. We also impose a cutoff at $z=4$. To carry out  a more realistic analysis, the photometric redshift error is also taken into account  by using a gaussian modulation with variable variance $\sigma_{z} = (1+z)\delta_{z} $. The $\delta_{ z}$ takes two different values, namely  0.05 and 0.01 in our analysis. The $\delta_z = 0.05$ corresponds to ongoing projects and $\delta_z = 0.01$ is assumed for high precision photometric redshift estimation.

The shear angular power spectrum is defined in the flat sky Limber approximation \citep{loverde2008extended} $\langle \gamma_{\ell}^{\rm A}(i) \gamma_{\ell^{\prime}}^{\rm B}(j) \rangle = (2\pi)^2 \delta_{\ell \ell^{\prime}}^D C_\ell^{\rm A B} (i,j)$, where $\gamma_{\ell}^{\rm A}(i)$ is the Fourier transform of $\gamma^{\rm A}_{\theta}(i)$. Therefore, the result takes the following form:
\begin{equation}\label{eq:ab}
C_\ell^{\rm A B} (i,j)=\frac{9}{4}(\Omega_m H_0^2)^2 \int \mathrm{d}\chi \frac{W_i^{\rm A}W_j^{\rm B} }{\chi^2}\mathcal{P}^{\rm A B}\left(\frac{\ell}{\chi},z\right)
\end{equation}
where we have used the definition of 3D shear field power spectra: $\langle \gamma^{\rm A}\left(\bold{k},z\right) \gamma^{\rm B}\left(\bold{k}^{\prime},z\right) \rangle =\frac{9}{4}(\Omega_m H_0^2)^2 (2\pi)^3 \delta^D(\bold{k}-\bold{k}^{\prime})\mathcal{P}^{\rm A B}(k,z)$. Here the angle bracket indicates ensemble average and $\rm A$ and $\rm B$ indices could be replaced by both G and I. {Throughout this analysis, we use linear alignments model as a theoretically simple model whereas it matches with observation \citep{catelan2001intrinsic,hirata2004intrinsic}. Therefore, we have for 3D power spectra of WL and IA shear field:
\begin{eqnarray}
     {\mathcal P}^{\rm GG} \left(k,z\right)  &=& {\mathcal P}^{\rm nl}_{\delta_m\delta_m}\left(k,z\right) \\
     {\mathcal P}^{\rm II} \left(k\right)  &=&{\mathcal P}_{\delta_m\delta_m}\left(k\right) \\
     {\mathcal P}^{\rm GI}\left(k,z\right)  &=& {\mathcal P}^{\rm IG}\left(k,z\right) = \sqrt{{\mathcal P}_{\delta_m\delta_m}\left(k\right) {\mathcal P}^{\rm nl}_{\delta_m\delta_m}\left(k,z\right)} \nonumber\\
     &&
\end{eqnarray}
the superscript "$\rm nl$" refers to non-linear matter power spectrum. $\delta_m({\bf k})$ is the linear matter density contrast which is evolved through the linear growth \citep{linder2003cosmic}. In this paper we use CAMB code \footnote{\texttt{CAMB (http://camb.info)}} to compute present and non-linear mater power spectrum \citep{Lewis:1999bs}. We also consider a shot noise term added to the full cosmic shear angular power spectrum:
\begin{equation}\label{eq:datavector}
N_{\ell}(i,j) = \frac{\sigma_{\epsilon}^2}{\bar{n}}\delta_{ij}^D
\end{equation}
the $\sigma_{\epsilon} $ is gaussian ellipticity dispersion and $\bar{n}$ is the average number of galaxy in each bin. We use the Euclid like surveys specification \citep{amendola2018cosmology} to determine the appropriate values. Subsequently, we set $\sigma_{\epsilon} = 0.25$, $f_{\rm sky}=0.364$ and $n_{\rm galaxy} = 30 \ [\rm arcmin^{-2}]$. We also change the $\sigma_{\epsilon}$ to 0.15 to evaluate the result for DES like surveys.

Finally the shear angular power spectrum have following parts:
\begin{equation}\label{eq:angulpospec}
\{C_\ell^{\rm AB}(i,j)\}:\{C_\ell^{\rm GG}(i,j), \ C_\ell^{\rm II}(i,j), \ C_\ell^{\rm IG}(i,j), \ C_\ell^{\rm GI}(i,j)\}
\end{equation}
where GG, II and either IG or GI correspond to WL, IA auto- and their cross-correlations, respectively.  Therefore, the full cosmic shear angular power spectrum for $(i,j)$ bin pair is given by:
\begin{eqnarray}\label{eq:cs}
C_{\ell}^{\rm I+G}(i,j)&=&C_\ell^{\rm GG}(i,j)+ C_\ell^{\rm II}(i,j)\nonumber\\
&&+\left(1-\frac{\delta_{ij}}{2}\right) \left[\ C_\ell^{\rm IG}(i,j)+\ C_\ell^{\rm GI}(i,j)\right]
\end{eqnarray}

Now, we turn to estimate the effect of SSC on the IA and full cosmic shear. To this end, we define the observed data vector for full cosmic shear angular power spectrum by $[ \tilde{C}_{\ell}^{\rm I+G}\left( i,j \right) ]_{\mu}$ and $\mu$ runs from one to  $\begin{pmatrix}M+1\\2\end{pmatrix} \times \ell$  for $M$ bins. The observed data vector is computed using following estimator:
\begin{equation}
\tilde{C}_{\ell}^{\rm I+G} (i,j)= \frac{1}{4\pi f_{\rm sky}} \int _{|\bm{\ell}|\in \ell}\frac{\mathrm{d}^2{\bm{\ell}}}{2\pi \ell \Delta \ell}  \tilde{\gamma}_{\bm{\ell}}^{\rm I+G}(i) \tilde{\gamma}_{-\bm{\ell}}^{\rm I+G}(j)
\end{equation}
We set the logarithmic equal spaced, $\Delta \log \ell = 0.12$ throughout this paper. The $\tilde{\gamma}_{{\bm{\ell}}}^{\rm A}(i)$ is related to the theoretical cosmic shear by convolutional integral over window of survey:
\begin{equation}
\tilde{\gamma}_{\bm{\ell}}^{\rm A}(i) = \int \frac{\mathrm{d^2}{\bm{\ell}}^{\prime}}{(2\pi)^2}W_s\left({\bm{\ell}}^{\prime}\right) \gamma_{{\bm{\ell}} - {\bm{\ell}}^{\prime}}^{\rm A}(i)
\end{equation}
the $W_s\left({\bm{\ell}} \right)$ is survey window and for a partial sky disk shape coverage, we have:
\begin{equation}
W_s\left( \ell \right) = 8\pi f_{\rm sky} \frac{J_1\left(\ell r_0 \right)}{\ell r_0}
\end{equation}
the $J_1(x)$ is ordinary Bessel function of first kind. The $r_0 = 2\sqrt{f_{\rm sky}}$ is effective radius of survey. Subsequently, observed data vector is $\left\langle\tilde{C}^{{\rm A} {\rm B}}_\ell (i,j) \right\rangle  = C^{{\rm A} {\rm B}}_\ell (i,j)$ and we have the associated covariance matrix, which is a sum of the 16 terms as follows:
\begin{multline}
 \bold{C}\left( \tilde{C}_{\ell}^{\rm I+G} (i,j) , \tilde{C}_{\ell^{\prime}}^{\rm I+G}(p,q)  \right) = \\ \sum_{\rm ABCD} \left[\left\langle \tilde{C}_{\ell}^{\rm A B} (i,j) \tilde{C}_{\ell^{\prime}}^{\rm C D}(p,q)  \right\rangle - \left\langle \tilde{C}_{\ell}^{\rm A B} (i,j)\right\rangle  \left\langle \tilde{C}_{\ell^{\prime}}^{\rm C D}(p,q)\right\rangle\right]
\end{multline}
\begin{figure}
\centering
  \includegraphics[width=1.\columnwidth]{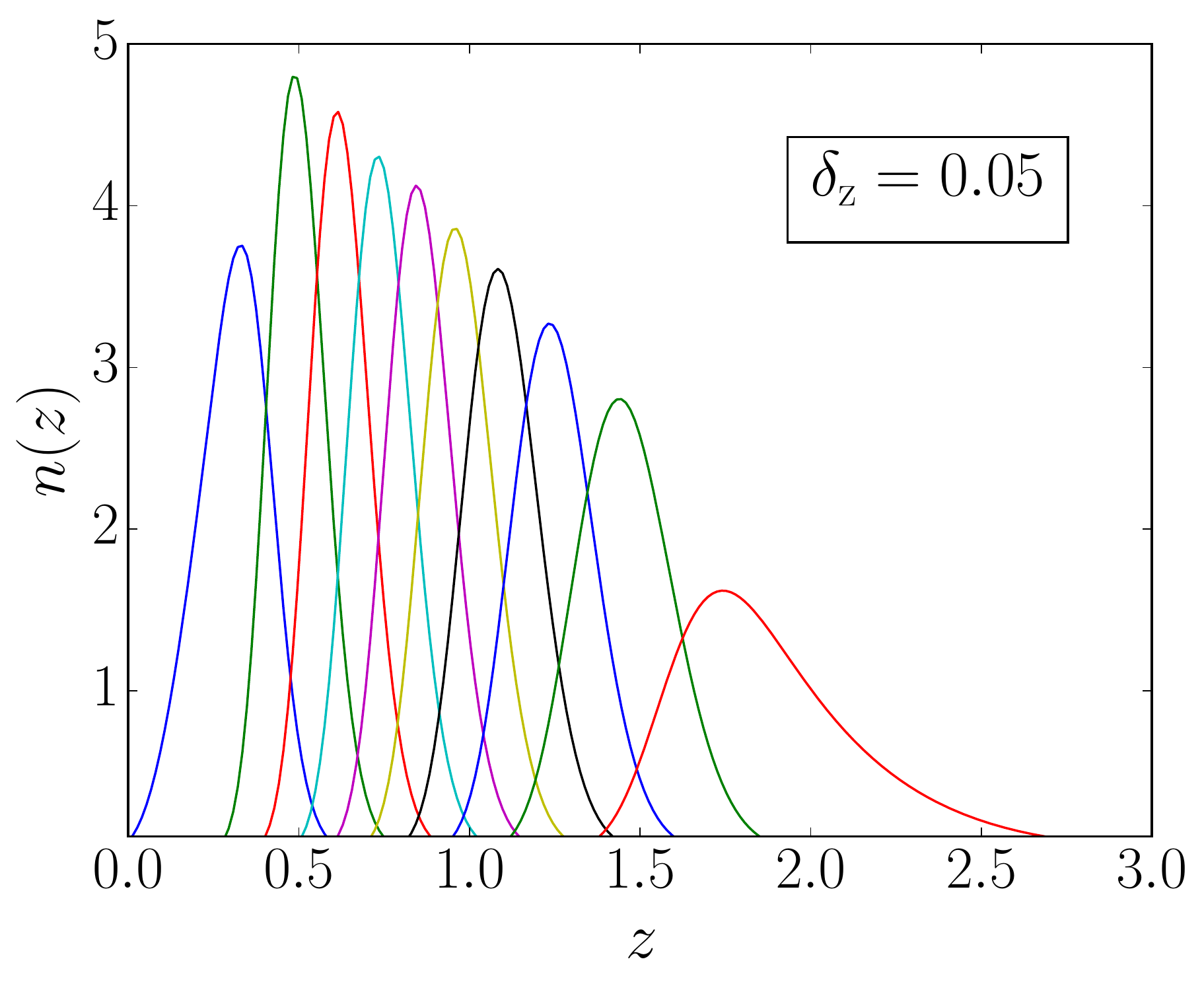}
  \caption{Color online: Normalized galaxy distribution with equally divided 10 redshift bins specified by different colored lines. The $\delta_{\rm z} = 0.05$ representing the photometric redshift error.}
  \label{fig:nz}
\end{figure}

The covariance matrix has in principle three components, the Gaussian covariance (basic), the connected non-Gaussian part and the super sample covariance. It is wroth noting that the contribution of various part on the total covariance matrix depends on different characteristics, however, the contribution of connected non-Gaussian covariance which is related to corresponding trispectrum
\citep{scoccimarro1999power,cooray2001power,takada4170mnras395,bertolini1604jcap6,bertolini2016non,barreira2017responses,barreira1705jcap11,barreira2018complete,takahashi2019covariances}, is  not significant \citep{barreira2018complete}. Subsequently, we turn to compute the basic and SSC components and we have $ \bold{C} = \bold{C}_{\rm basic} + \bold{C}_{\rm SSC}$. The elements of the first term in $\bold{C}$ containing the Gaussian part is given by:
\begin{multline}\label{eq:G covamatrix}
\bold{C}_{\rm basic}\left(  \tilde{C}_{\ell}^{\rm I+G} (i,j) , \tilde{C}_{\ell^{\prime}}^{\rm I+G}(p,q)  \right) =\frac{\delta_{\ell\ell'}}{2f_{\rm sky}\ell \Delta \ell}  \\
 \sum_{\rm ABCD}\left[C^{{\rm A} {\rm C}}_\ell(i,p)C^{{\rm B} {\rm D}}_\ell(j,q)+C^{{\rm A} {\rm D}}_\ell(i,q) C^{{\rm B} {\rm C}}_\ell(j,p)\right]
\end{multline}
Also, the $\bold{C}_{\rm SSC}$ which is generally non-diagonal is represented by \citep{barreira2018complete,wadekar2019galaxy}:
\begin{multline}\label{eq:SSC covamatrix}
\bold{C}_{\rm SSC} \left(  \tilde{C}_{\ell}^{\rm I+G} (i,j) , \tilde{C}_{\ell^{\prime}}^{\rm I+G}(p,q)   \right) =  \\
 \frac{81}{16}(\Omega_m H_0^2)^4 \sum_{\rm ABCD}  \int \mathrm{d}\chi \frac{W^{\rm A}_i(\chi)W^{\rm B}_j(\chi)W^{\rm C}_p(\chi)W^{\rm D}_q(\chi)}{\chi^6} \\\times \left(\frac{68}{21}\right)^2 \sigma^2 \left( \chi \right) {\mathcal P}^{\rm AB}\left(\frac{\ell}{\chi},z\right) {\mathcal P}^{\rm CD}\left(\frac{\ell^{\prime}}{\chi},z\right)
\end{multline}
where the $\sigma^2(\chi)$ is the variance of matter in the survey area:
\begin{equation}
\sigma^2(\chi) = \frac{1}{(4\pi f_{\rm sky})^2} \int \frac{\ell \mathrm{d} \ell}{2\pi} |W_s(r_0 \ell)|^2  \mathcal{P}_{\delta_m \delta_m}\left(\frac{\ell}{\chi}\right)
\end{equation}
The effect of SSC on the power spectrum in the weakly non-linear regime for unbiased tracer (the last part in Eq. (\ref{eq:SSC covamatrix})) has generally two terms of the large scale density contrast and the large scale tides, respectively \citep{akitsu2017large}.  We ignore both the large scale tides \citep{akitsu2018impact} and the $k$-dependent part for the large scale density contrast term in the tree level perturbation theory \citep{bernardeau2002large} which is not very significant for the range $k\lesssim 0.4$ [$h$/Mpc] \citep{takada2013power},  consequently, we achieve  $(68/21)^2$ coefficient appeared in the last part of Eq. (\ref{eq:SSC covamatrix})). It is wroth noting that, we also disregard the effect of connected non-Gaussian terms compared to large scale density contrast terms \citep{barreira2018complete}.
Finally, we compute the cumulative signal to noise ratio (SNR) with and without considering the SSC effect:
\begin{equation}
{ \left( \rm \frac{S}{N} \right)_{\star} } \equiv \sqrt{[C_{\ell}^{\rm I+G} (i,j) ]_{\mu} [\bold{C}_{\star}^{-1}]_{\mu \nu} [C_{\ell}^{\rm I+G}(i,j)]_{\nu}}
\end{equation}
Here the subscript $\star$ refers to the different  type of cosmic shear covariance matrix with and  without corresponding SSC.
We consider the full treatment as $\gamma_{\rm SSC}^{\rm I+G}$ and in order to show the contribution of including SSC for IA, we define $\gamma_{\rm semi}^{\rm I+G}$, in which only the SSC term of WL has been included in the full cosmic shear.
\begin{figure*}
\centering
  \includegraphics[width=1.5\columnwidth]{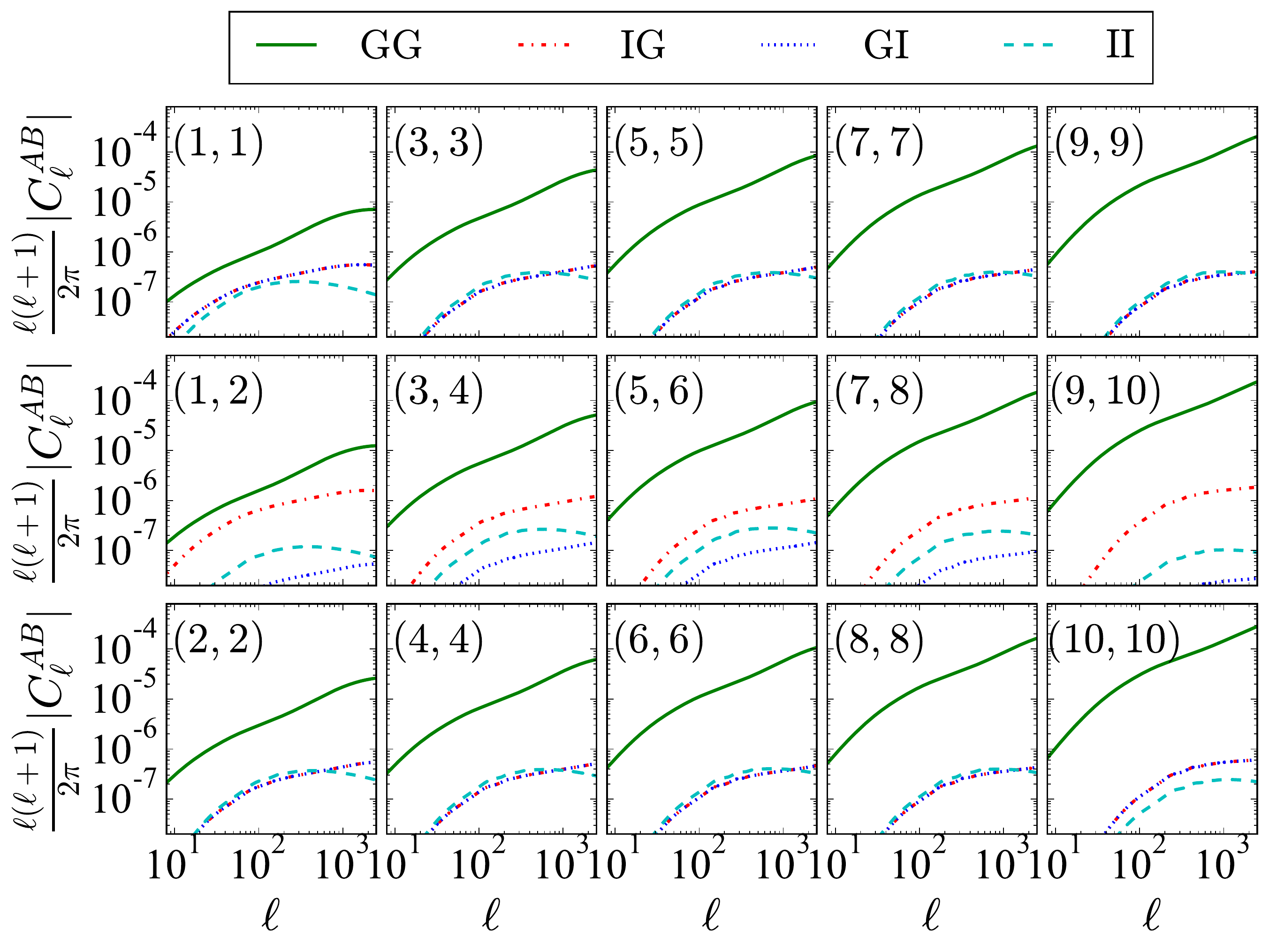}
  \caption{Absolute value for various components of full cosmic shear angular power spectrum in different combinations of redshift bins $\{(i,i),(i,i+1),(i+1,i+1)\}$. Green solid line stands for $C_\ell^{\rm GG}$, cyan dashed lines, red dot-dashed lines and blue dotted lines correspond to $C_{\ell}^{\rm II}$, $C_{\ell}^{\rm IG}$ and $C_{\ell}^{\rm GI}$, respectively. Here we have used the value $A_{\rm I}$ = 1.0 for amplitude of intrinsic alignment.}
  \label{fig:cl}
\end{figure*}

To realize what statistical changes happen by including and excluding the SSC of $\gamma^{\rm I}$ when two cases $\gamma^{\rm I+G}_{\rm SSC}$ and  $\gamma^{\rm I+G}_{\rm semi}$ to be taken into account according and to examine the error propagation into the precise evaluation of cosmological parameters, we utilize the Fisher information matrix as \citep{1997ApJ...480...22T}:
\begin{align}
    \bold{F}_{\mu\nu} = \frac{1}{2} {\rm Tr}\left[\bold{C}^{-1} \frac{\partial_\mu \bold{C}}{\partial \Theta_{\mu}}  \bold{C}^{-1} \frac{\partial_\nu \bold{C}}{\partial \Theta_{\nu}}\right]
\end{align}
Now, we should compute the  inverse and derivative of covariance matrix with respect to underlying cosmological parameters, numerically.
As an illustration, the shape of $\bold{C}$ in two redshift bins in the matrix form is:
\begin{widetext}
\begin{multline}
\bold{C} = \bold{C}_{\rm basic}+
\begin{pmatrix}
\bold{C}_{\rm SSC}\left(\tilde{C}^{\rm I+G}_2(1,1),\tilde{C}^{\rm I+G}_2(1,1) \right) & \bold{C}_{\rm SSC}\left(\tilde{C}^{\rm I+G}_2(1,1),\tilde{C}^{\rm I+G}_2(1,2) \right)    &\cdots &  \bold{C}_{\rm SSC}\left(\tilde{C}^{\rm I+G}_2(1,1),\tilde{C}^{\rm I+G}_{\ell}(2,2) \right)\\
\bold{C}_{\rm SSC}\left(\tilde{C}^{\rm I+G}_2(1,2),\tilde{C}^{\rm I+G}_2(1,1) \right)& \bold{C}_{\rm SSC}\left(\tilde{C}^{\rm I+G}_2(1,2),\tilde{C}^{\rm I+G}_2(1,2) \right)   &   \cdots &  \bold{C}_{\rm SSC}\left(\tilde{C}^{\rm I+G}_2(1,2),\tilde{C}^{\rm I+G}_{\ell}(2,2) \right)\\
\vdots  & \vdots  & \vdots  \\
\bold{C}_{\rm SSC}\left(\tilde{C}^{\rm I+G}_{\ell}(2,2),\tilde{C}^{\rm I+G}_2(1,1) \right) &\bold{C}_{\rm SSC}\left(\tilde{C}^{\rm I+G}_{\ell}(2,2),\tilde{C}^{\rm I+G}_2(2,1) \right)  & \cdots & \bold{C}_{\rm SSC}\left(\tilde{C}^{\rm I+G}_{\ell}(2,2),\tilde{C}^{\rm I+G}_{\ell}(2,2) \right)
\end{pmatrix}
\end{multline}
\end{widetext}
Where the elements will be given by a sum over Eq. (\ref{eq:G covamatrix}) and Eq. (\ref{eq:SSC covamatrix}).
In the next section, we will give our main results concentrating on the contribution of SSC in the cosmic shear observation.

\begin{figure*}
\centering
  \includegraphics[width=1.5\columnwidth]{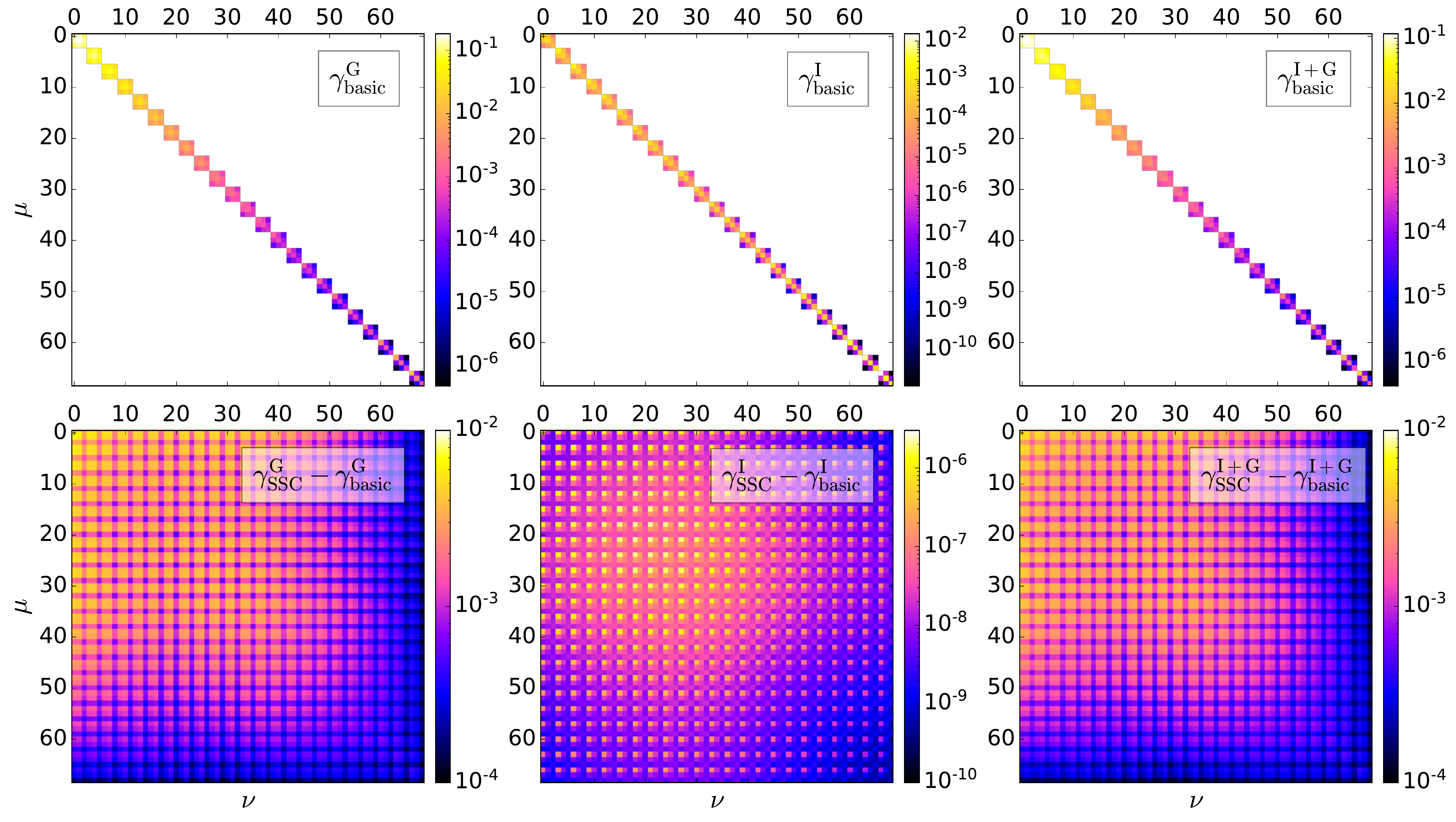}
  \caption{Contributions of different component in full cosmic shear covariance matrix. Basic block-diagonal part (upper panel) is compared to SSC contribution (lower panel). The color-bar represent the $\bold{C}_{\mu \nu}/([ C_{\ell}^{\rm I+G}(i,j) ]_{\mu} \times [ C_{\ell^{\prime}}^{\rm I+G} (p,q) ]_{\nu})$, values of covariance matrix which is normalized to corresponding angular power spectrum in order to better visualization. Here we consider redshift bin combination $(1,2)$.}
  \label{fig:cov1}
\end{figure*}

\section{Results}\label{sec:results}
As discussed in previous section, we consider 10 redshift bins with equal fraction of total galaxies in each bin. In order to carry out the Fisher forecast analysis with less computational cost, we consider combination of each two neighbor bins. The first ($z\in[0.0-0.42]$) and the second ($z\in[0.42-0.56]$) redshift bins among 10 bins are used to show the effect of SSC when it is included for IA cosmic shear. We also use the other 4 combination $(3,4)$, $(5,6)$, $(7,8)$ and $(9,10)$ to compare the results to show the importance of this effect by going to deep field regime. As indicated in Fig. \ref{fig:nz}, the redshift photometry error is applied to build normalized galaxy distribution in each bin.

Different components of full cosmic shear angular power spectrum have been illustrated in Fig. \ref{fig:cl}. Increasing the number of bins with fixed redshift range leads to decrease the width of each bin, consequently, the $\gamma^{\rm G}$ contribution in IG terms becomes weak in a single redshift bin. For all cases, the gravitational lensing plays dominant role  compared to other portions. In addition, the angular power spectrum of the $\gamma^{\rm G}$ term  possesses the higher value for $i\neq j$, due to increasing the spanned redshift range. The II term is effectively very small for cross-correlation of different redshift bins. The GI terms are same as IG in the same redshift bins. However for a different redshift bin analysis labeled  by $(i,j)$ an for the case $i<j$,  the GI term goes to zero by increasing the precision in photometric redshift estimation. One can manipulate the contribution of cross term in the power spectrum by changing $\delta_z$.

The corresponding covariance matrix of different components of full cosmic shear are indicated in Fig. \ref{fig:cov1}. For better illustration, we divide the covariance matrix $\bold{C}_{\mu \nu}$ by $[ C_{\ell}^{\rm I+G}(i,j) ]_{\mu} \times [ C_{\ell^{\prime}}^{\rm I+G} (p,q) ]_{\nu}$ vectors. As indicated in the upper panels of Fig. \ref{fig:cov1}, the G-, I- and their full contribution in the $\gamma_{\rm basic}$ are all $3 \times 3 $ block diagonal matrix. The SSC contribution of each covariance matrix are shown in the lower panels of Fig. \ref{fig:cov1}. The extension in the range of values in the mentioned matrices are due to extra bin cross terms. The $\gamma_{\rm basic}^{\rm I}$ has maximum impact at low $\ell$ while the main contribution of SSC due to intrinsic alignments, notified by  $\gamma_{\rm SSC}^{\rm I} - \gamma_{\rm basic}^{\rm I}$, comes from almost middle $\ell$. When we add the intrinsic alignments to the basic covariance matrix, the SNR decrease at low $\ell$ while it asymptotically goes toward  $\gamma^{\rm G}_{\rm basic}$ at high $\ell$. The SSC however, becomes more significant by accumulating more $\ell$.  We compute the SNR in redshift bins $(1,2)$ illustrated in the upper panel of Fig. \ref{fig:sn}. Including SSC for WL cosmic shear, decreases SNR by  $\approx 50\%$ with respect to $\gamma^{\rm G}_{\rm basic}$, for large multipole. Our results demonstrate that including the intrinsic alignments shear to gravitational lensing,  the SNR is reduced for low $\ell$.  Adding the intrinsic alignments and considering the associated SSC-effect moderate the SNR by $\approx 5\%$ by each of them (lower panel of Fig. \ref{fig:sn}).

In order to trace the dependency of intrinsic alignments SSC on the survey characteristics and sample properties, we compute the SNR for four combinations of  two different ellipticity dispersions and intrinsic alignments amplitude at redshift bins $(1,2)$. As indicated in Fig. \ref{fig:sn2}, by decreasing the mean ellipticity dispersion from 0.25 to 0.15, the overall SNR increases. As depicted by solid and dotted lines in Fig. \ref{fig:sn2}, the SNR is reduced by almost 10\% and 20\% for $\sigma_{\epsilon}=0.25$ and $\sigma_{\epsilon}$, respectively, when we assume the SSC-effect of IA in a full cosmic shear analysis. The SSC on IA will be more significant if we consider a sample with high value of intrinsic alignments amplitude. in this case, the relative SNR with respect to $\gamma^{\rm I+G}_{\rm semi}$ is declined more than $\approx 20\%$ for $\sigma_{\epsilon}=0.25$ and around $\approx 40\%$ for $\sigma_{\epsilon}=0.15$.

To examine the effect of redshift bin, we compute SNR for various bin combinations in Fig. \ref{fig:sn3}. To obtain a magnified effect, we take extreme values for relevant parameters as $A_{\rm I}=3$, $\delta_z=0.01$ and $\sigma_{\epsilon}=0.15$. Going to high redshift bin, the SNR grows for high $\ell$.
 Our results demonstrate that, the SSC of IA has almost $\approx 15\%$ and $\approx 5\%$ contribution for $(3,4)$ and $(5,6)$, redshift bins, respectively, for our selected parameters,  subsequently, the lower redshift bin, the higher impact for including SSC of IA.

\begin{figure*}
\centering
  \includegraphics[width=1.2\columnwidth]{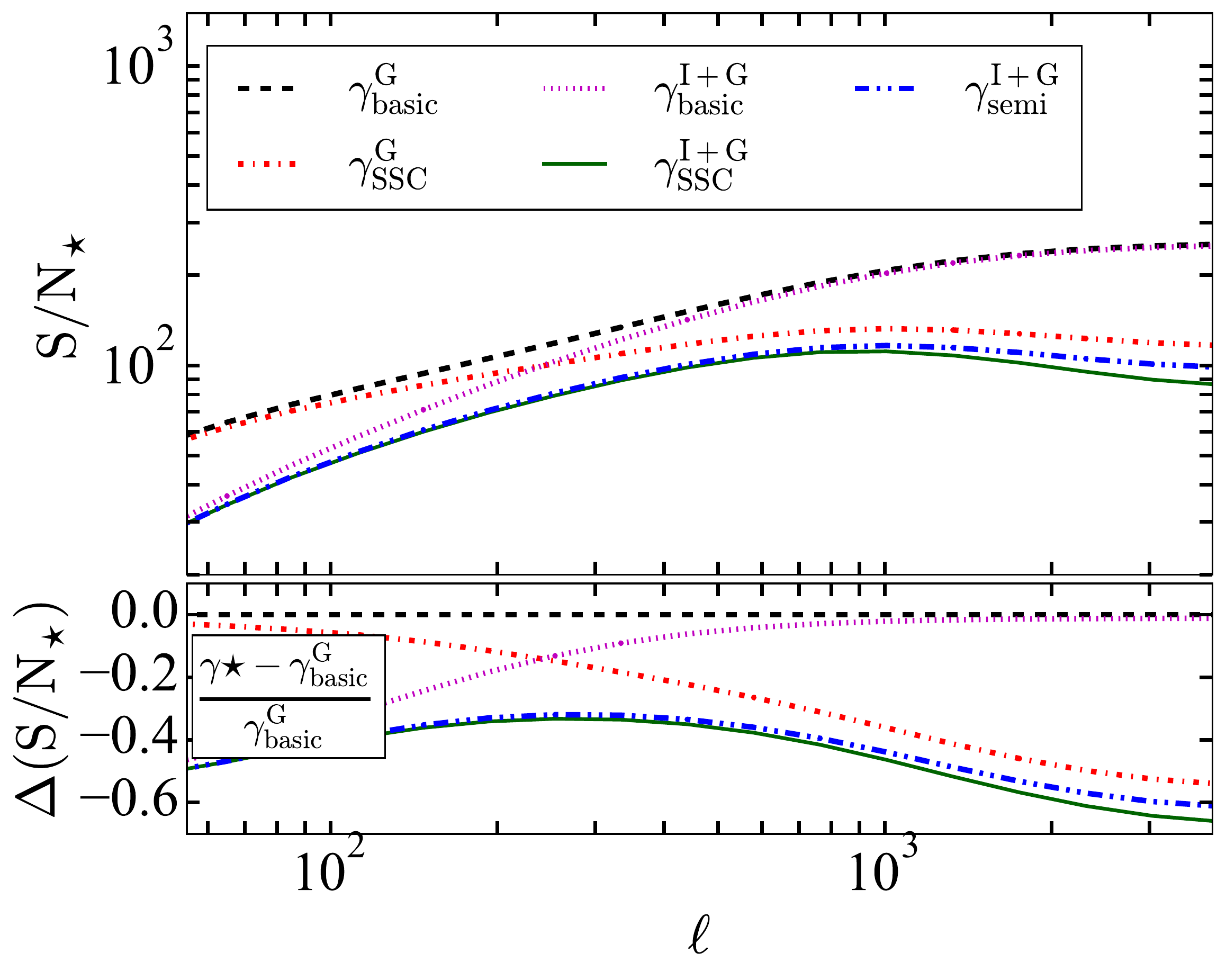}
  \caption{Upper panel corresponds to the cumulative signal to noise ratio for different components of full cosmic shear signal. The basic (black dash line) and SSC included (red dot dash line) for cosmic shear.  The purple dot line and green solid line are devoted to basic and SSC included full cosmic shear for redshift bin combination $(1,2)$, respectively. For the full cosmic shear signal when only the SSC-effect on the gravitational lensing to be taken into account and specified by subscript "semi" is represented by blue dash dot-dot line. Lower panel illustrates the relative SNR with respect to the basic cosmic shear. Here we set the parameters according to Euclid like survey, $\delta_z=0.05$ and $\sigma_{\epsilon}=0.25$ and for our case, we took $A_{\rm I}=1.0$.}
  \label{fig:sn}
\end{figure*}

\begin{figure*}
\centering
 \includegraphics[width=1.5\columnwidth]{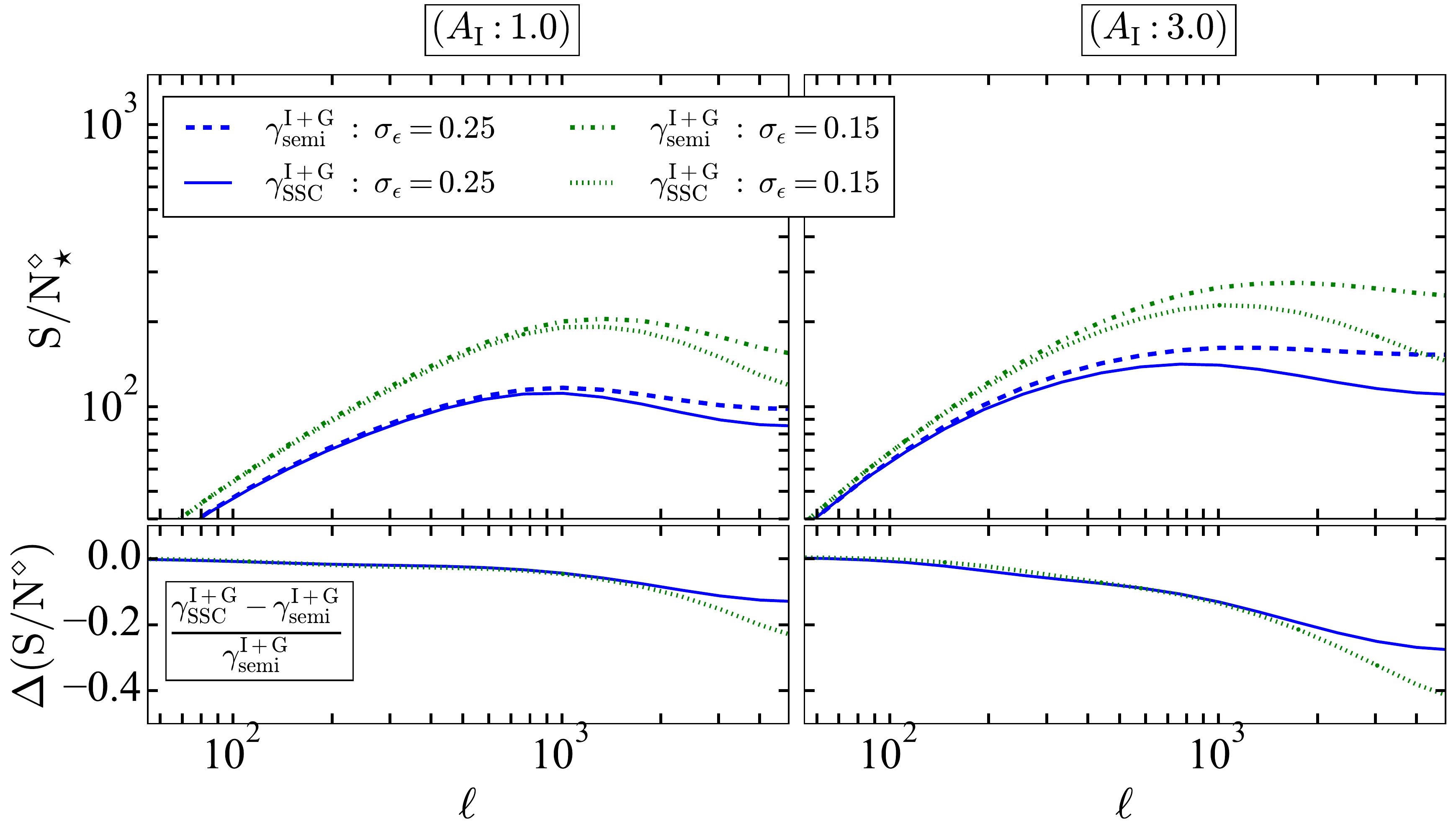}
  \caption{ Upper panel indicates the full and semi cosmic shear computed for $\sigma_{\epsilon}=0.25$ and $\sigma_{\epsilon}=0.15$. The left part is for intrinsic alignments amplitude $A_{\rm}=1.0$ while the right part illustrates the results for $A_{\rm I}=3.0$. Lower panel corresponds to relative cumulative signal to noise ratio of full cosmic shear with respect to semi cosmic shear for $\sigma_{\epsilon} = 0.25$ (blue solid line) and $\sigma_{\epsilon} = 0.15$ (green dot line). Here we consider redshift bin combination $(1,2)$.}
  \label{fig:sn2}
\end{figure*}

\begin{figure*}
\centering
 \includegraphics[width=1.5\columnwidth]{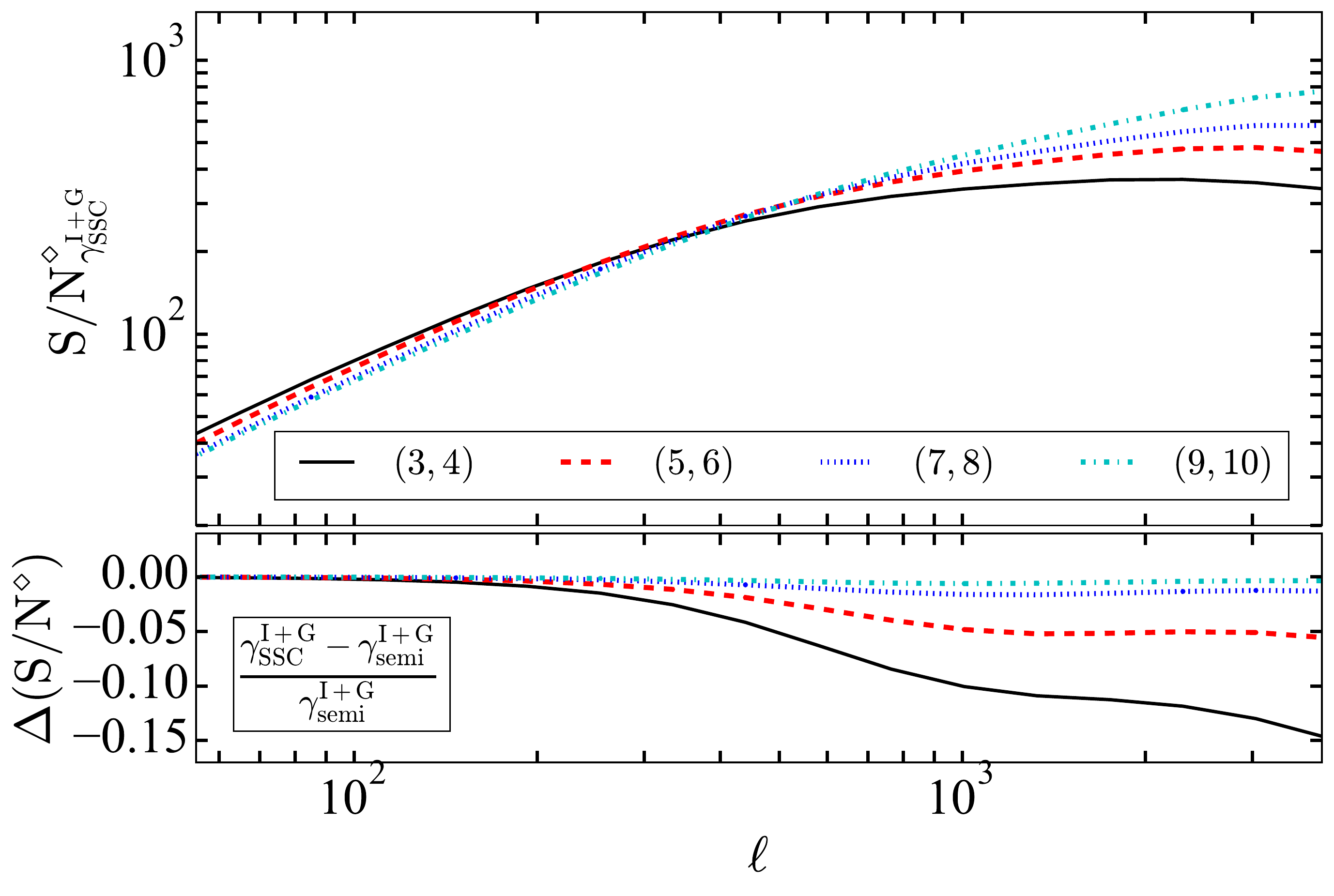}
  \caption{Upper panel corresponds to SNR of full cosmic shear signal for different redshift bins, $(3,4)$ (black solid line), $(5,6)$ (red dash line), $(7,8)$ (blue dot line) and $(9,10)$ (green dash dot line). Lower panel illustrates the relative cumulative signal to noise ratio with respect to the case, where the SSC has only been considered for gravitational lensing, known as "semi". We took, $A_{\rm I}=3$, $\delta_z=0.01$ and $\sigma_{\epsilon}=0.15$.   }
  \label{fig:sn3}
\end{figure*}

\begin{figure*}
\centering
  \includegraphics[width=1.5\columnwidth]{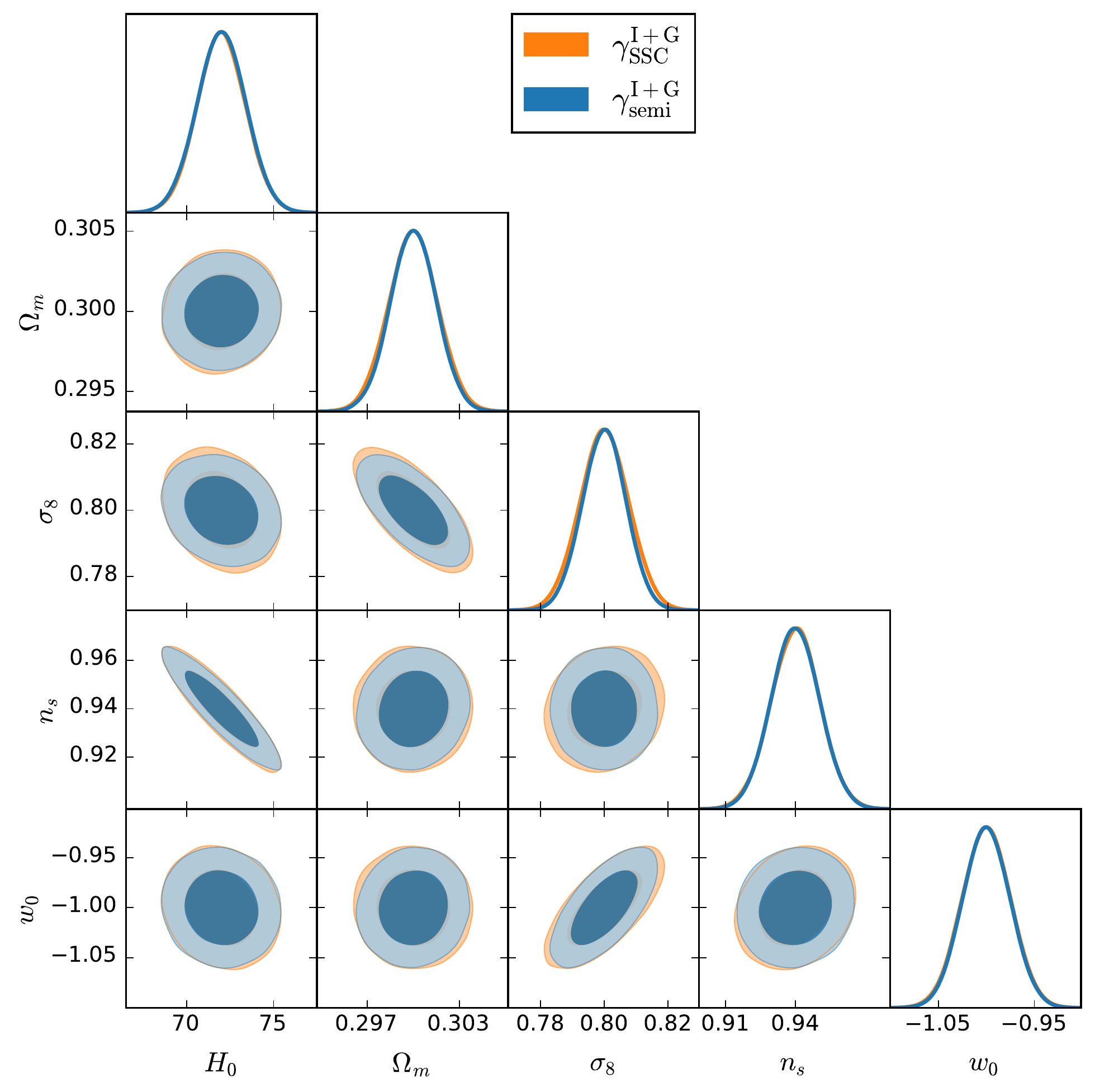}
  \caption{The one and two dimensional marginalized likelihood functions of cosmological parameters derived by Fisher forecast analysis for full cosmic shear signal (orange) compared to the semi case (blue). Here we consider redshift bin combination $(1,2)$. We took, $A_{\rm I}=3$, $\delta_z=0.01$ and $\sigma_{\epsilon}=0.15$. For each case, the dark color corresponds to $1\sigma$ confidence interval and light color shows the $2\sigma$ region. In this case, two contour sets are almost identical. }
  \label{fig:final}
\end{figure*}

\begin {table*}
\begin{center}
\caption {The 68\% confidence interval for cosmological parameters computed by Fisher forecast analysis in the context of cosmic shear observation with and without SSC effect. We computed the error values for two combinations of redshift bins, namely  (1,2) and (2,3).} \label{tab:1}
\end{center}
\begin{center}
\begin{tabular} { c c c c c c c c c }
\hline
Amplitude  & \multicolumn{ 4 }{c}{$A_{\rm I}=1$} & \multicolumn{ 4 }{c}{$A_{\rm I}=3$} \\
\hline
Bins       &   \multicolumn{ 2 }{c}{(1,2)}    & \multicolumn{ 2 }{c}{(2,3)}  &   \multicolumn{ 2 }{c}{(1,2)}    & \multicolumn{ 2 }{c}{(2,3)} \\
\hline
Type  &  $\gamma^{\rm I+G}_{\rm semi}$ &  $\gamma^{\rm I+G}_{\rm SSC}$&  $\gamma^{\rm I+G}_{\rm semi}$ &  $\gamma^{\rm I+G}_{\rm SSC}$&  $\gamma^{\rm I+G}_{\rm semi}$ &  $\gamma^{\rm I+G}_{\rm SSC}$ &  $\gamma^{\rm I+G}_{\rm semi}$ & $\gamma^{\rm I+G}_{\rm SSC}$ \\
\hline
$\Delta_{\Omega_m} $& $ 0.0030 $ & $ 0.0030 $ & $ 0.0025 $ & $ 0.0025 $ & $ 0.0015 $ & $ 0.0016 $ & $ 0.0018 $ & $ 0.0018 $\\
$\Delta_{\sigma_8} $   & $ 0.0079 $ & $ 0.0080 $ & $ 0.010 $   & $ 0.010 $   & $ 0.0069 $ & $ 0.0077 $ & $ 0.010  $  & $ 0.010 $ \\
$\Delta_{H_0}$            & $ 1.5$        & $ 1.5 $       & $ 1.2 $       & $ 1.2 $       & $ 1.4 $       & $ 1.4 $       & $ 1.1 $       & $ 1.1 $      \\
$\Delta_{n_s}$             & $ 0.015$    & $ 0.016$    & $ 0.009  $  & $ 0.009 $   & $ 0.010 $   & $ 0.011 $   & $ 0.008 $   & $ 0.008  $ \\
$\Delta_{w_0}$            & $ 0.029$    & $0.029$     & $ 0.027 $   & $ 0.027 $   & $ 0.025 $   & $ 0.025 $   & $ 0.030 $   & $ 0.030 $  \\
\hline
\end{tabular}
\end{center}
\end{table*}

We perform Fisher formalism for forecasting cosmological parameter constraints in the context of $\gamma^{\rm G}$, $\gamma^{\rm I}$ and $\gamma^{\rm I+G}$. We compare our results for two cases, namely for the first case, we include only the SSC-effect on the gravitational lensing, while for the second case, we compute the super sample covariance of the full cosmic shear. Accordingly, we are readily able to examine the contribution of finite size of surveys in the cosmology of weakly non-linear regime. In Fig. \ref{fig:final}, we illustrate the marginalized relative likelihood function and contour plots for the $\gamma^{\rm I+G}_{\rm semi}$ (dark blue) and $\gamma^{\rm I+G}_{\rm SSC}$ (light orange)  for redshift bin $(1,2)$. As shown in this plot, in the presence of shot noise and photometric redshift error, the SSC of IA has almost no effect on the confidence interval of cosmological parameters and meanwhile, the $\sigma_8$ experiences a marginal change.  To make more sense, we define $\Delta_{\diamond}$ as the associated error-bar of the underlying cosmological parameters (the "$\diamond$" symbol is replaced by each cosmological parameters considered in this paper).  In Table \ref{tab:1}, we report $\Delta_{\diamond}$ computed by Fisher forecast approach in the context of cosmic shear observations due to SSC effect at $1\sigma$ confidence interval. By increasing the redshift bin, the difference between uncertainties in cosmological parameters with (labeled by "SSC") and without (labeled by "semi") SSC-effect of IA decreases.
\section{Summary and Conclusions} \label{sec:summary}
In this paper, we examined the power spectrum of cosmic shear signal including intrinsic alignments (IA) and weak gravitational lensing (WL) to evaluate the cosmological parameters spaces when the supper sample covariance to be taken into account. To this end, we utilized the Fisher forecast formalism. This approach enables us to determine the size of parameters space and  how the correlations between parameters to be affected by the $beat$ $coupling$ phenomenon.

The contribution of various parts in the cosmic shear signal have been investigated and our results demonstrated that, the angular power spectrum of the $\gamma^{\rm G}$ term has dominant role compared to II part and its value reach to almost  higher value for $i\neq j$, due to increasing the spanned redshift range. Increasing the precision in photometric redshift estimation enables us to manipulate the contribution of cross term in the power spectrum (Fig. \ref{fig:cl}). Decreasing the mean ellipticity dispersion leads to increase SNR of full cosmic shear signal. Utilizing  the higher value for intrinsic alignments amplitude  and the lower value for $\sigma_{\epsilon}$ enhance the reduction of SNR for high $\ell$ when the SSC-effect on the IA is added to the rest part.

Our analysis confirmed that, contribution of SSC effect plays role almost for intermediate and high $\ell$ (Fig. \ref{fig:sn}). To achieve more precise evaluation, by adding the intrinsic alignments shear to gravitational lensing, the SNR is reduced for low $\ell$. In principle,  the SSC contribution makes a deficit at high $\ell$. Meanwhile, the SSC-effect on the gravitational lensing keeps the main contribution and incorporating the intrinsic alignments itself and associated SSC-effect moderate the SNR by $\approx 5\%$ due to each components (lower panel of Fig. \ref{fig:sn}). The lower ellipticity dispersion and higher intrinsic alignment amplitude essentially imply that the IA SSC-effect should be included to achieve more robust cosmological results (Fig. \ref{fig:sn2}). The SSC-effect on the IA has higher impact on the low redshift bin analysis (Fig. \ref{fig:sn3}).

Considering the gravitational lensing and intrinsic alignments in the cosmic shear survey and by including the {\it beat coupling} phenomenon for WL leads  the considerable part of cosmological information for parameter estimation. However, including the SSC for IA marginally alters the confidence interval of $\sigma_8$ which could be relevant for more precise measurement (Fig. \ref{fig:final}).
Finally, we summarize some of our main results:
\itemize
\item The main contribution of including IA cosmic shear on the covariance matrix elements is at low $\ell$.
\item The SSC has important impact at middle and high multipoles for computing the SNR. While the {\it beat coupling} for WL has dominant impact compared to corresponding effect for IA.
\item The SNR is affected by mean ellipticity dispersion and amplitude of intrinsic alignments. While, The photometric redshift error has no considerable impact on the relative cumulative signal to noise ratio.
\item The SSC has a considerable effect for low redshift bin. The relative SNR decrease significantly by going to higher redshift bin.
\item Including SSC for IA does not change significantly the parameter errors, while the $\sigma_8$ statistical error marginally grows up to $\approx 10\%$ at $1\sigma$ confidence interval for a sample with high amplitude of the intrinsic alignments, low shot noise and low mean ellipticity dispersion.  Considering all available redshift bins and possible values for relevant parameters in weakly non-linear regime demonstrate that, the SSC-effect of IA in parameter constraints can be ignored.

It could be interesting to consider deeply non-linear impact on the SSC effect of full cosmic shear, precisely. Also, taking into account more complicated models for intrinsic alignments causes to get deep insight on the super sample covariance. The SSC might have a different influence on other intrinsic alignments models such as spiral galaxies as those would be a phenomenon on very small angular scales and at high redshift.  The connected non-Gaussian term to obtain more accurate results could be further noticed. The large-scale tidal fields effect is another subject to explore left for future study.

\section*{Acknowledgements}
The authors are really  grateful to Bj{\"o}rn Malte Sch{\"a}fer  for his very useful and constructive comments. SA and SMSM appreciate the hospitality of Astronomy center in Heidelberg University  to the Zentrum f{\"u}r Astronomie der Universit{\"a}t Heidelberg, where some parts of this analysis have been done.

\section*{Data Availability}
The data underlying this article will be shared on reasonable request to the corresponding author.

\newpage
\bibliography{SSCbiblio}{}
\bibliographystyle{mnras}
\bsp
\label{lastpage}
\end{document}